# Universal nuclear focusing of confined electron spins

Sergej Markmann[1], Christian Reichl[2], Werner Wegscheider[2] and Gian Salis[1]

[1]*IBM Research-Zurich, Säumerstrasse 4, 8803 Rüschlikon, Switzerland*
[2] *Solid State Physics Laboratory, ETH Zurich, 8093 Zurich, Switzerland*

**Abstract**

For spin-based quantum computation in semiconductors, dephasing of electron spins by a fluctuating background of nuclear spins is a main obstacle. Here we show that this nuclear background can be precisely controlled in generic quantum dots by periodically exciting electron spins. We demonstrate this universal phenomenon in many-electron GaAs/AlGaAs quantum dot ensembles using optical pump-probe spectroscopy. A feedback mechanism between the saturable electron spin polarization and the nuclear system focuses the electron spin precession frequency into discrete spin modes. Employing such control of nuclear spin polarization, the electron spin lifetime within individual dots can surpass the limit of nuclear background fluctuations, thus substantially enhancing the spin coherence time. This opens the door to achieve long electron spin coherence times also in lithographically-defined many-electron systems that can be controlled in shape, size and position.

Semiconductor quantum dots (QDs) exhibit tunable atomic-like electronic states [1]. For this reason, QDs are referred to as artificial atoms and may serve as hosts for spin quantum bits (qubits), the main building block for spin-based quantum computation [2]. One of the core problems of spin qubits is an undesired hyperfine interaction of electron or hole spins with the nuclear environment of the semiconductor host material. This issue is known as the central spin problem [3-6]. Even though the size of a self-assembled QD is few tens of nm the confined electrons interact with $n = 10^4$-$10^6$ nuclear spins [7]. In this way the electron or hole spins experience an effective magnetic field $B_n$ (Overhauser field) which arises from the averaged nuclear magnetic moments. Due to the statistical nature of the nuclear polarization, the Overhauser field fluctuates (on a scale $\Delta B_n \propto \sqrt{n}$), thus giving rise to electron spin decoherence. Reduction of random nuclear background has been successfully demonstrated utilizing spin echo techniques in electrostatically defined QDs [8-12] as well as in self-assembled QDs [13-15] by electrical and optical means, respectively. Control of nuclear field fluctuations has also been demonstrated using continuous-wave laser excitation in self-assembled quantum dots [16-18]. Furthermore, it has been shown that electron spin decoherence, arising from nuclear spin fluctuations, can be drastically reduced in inhomogeneously broadened QD ensembles. This is realized by a self-synchronization of the electron spin precession with the repetition rate of a spin-exciting laser pulse train [19] which was termed spin mode-locking. Spin mode-locking offers the prospect to overcome hyperfine-induced spin dephasing but up to now has only been observed in singly charged self-assembled QDs.

Different explanations for how electron spin precession is synchronized to the repetition rate of spin excitation have been proposed [20-22], but the details are not well understood yet.

Here we report that spin mode-locking is a universal phenomenon that also occurs in many-electron GaAs/AlGaAs QDs of variable diameters of up to 1800 nm. From optical pump-probe measurements, we find that mode-locking is absent for depolarized nuclear spins and emerges slowly (time scale of seconds) once the laser excitation pulse train is turned on. This suggests that spin mode-locking arises from dynamical nuclear polarization (DNP). We develop a model and show that resonant spin amplification leads to a sharp decrease of DNP if the laser repetition period matches an integer multiple of the spin precession period, thus providing a feed-back mechanism that drives nuclear polarization to a precise value. This mechanism works for all systems where electron spin polarization saturates and is therefore not limited to single-electron quantum dots. This opens up new playgrounds for spin manipulation and control in material systems with nuclear background. Importantly, by this mechanism, the nuclear spin polarization becomes locked within a distribution that is much narrower than the typical low-frequency fluctuations of nuclear spins in quantum dots, demonstrating that long spin lifetimes can also be obtained in lithographically defined QDs where shape, spacing and positioning can be perfectly controlled. Such periodical excitation of the electron spins as a preparation step before starting a series of quantum operations could be a much simpler technique to enhance coherence times compared to dynamical decoupling schemes using spin echo pulses.

**Spin mode-locking**

Samples are obtained from an n-doped GaAs/AlGaAs quantum well grown by molecular beam epitaxy. Arrays of disk-shaped QDs with diameters between 400 nm and 1800 nm are defined by e-beam lithography and dry etching (see method section). Each dot contains between hundreds and several thousands of electrons depending on the dot size, as estimated from the quantum well carrier density (see supplementary material). Since the dot diameter is smaller than the spin-orbit length, spin dephasing due to spin-orbit coupling is drastically suppressed [23-25] compared to a 2D electron gas, as has been recently shown in wires [26-27]. Hyperfine interaction with fluctuating nuclear spin polarization becomes the dominant contribution to spin dephasing for small QD sizes and scales reciprocally with the QD diameter. In an intermediate regime, the sum of both dephasing contributions (spin-orbit and hyperfine) has a minimum, thus resulting in an electron spin system with long-lived electron spins. We use time resolve Kerr rotation measurements as described in Ref. [28] to study the electron spin dynamics of a QD ensemble. The spin polarization in the ensemble is excited with optical pump pulses propagating along the $z$-direction, perpendicular to the sample ($x$-$y$) plane. An external magnetic field **B** is applied along the $x$-direction. The circularly polarized pump pulses generate an electron spin polarization along the $z$-direction, which subsequently precesses about **B** [Fig. 1 (a)]. The spin polarization component $S_z$ is measured by a time-delayed linearly polarized probe pulse via the magneto-optical Kerr effect. The picosecond-long pulses arrive with a repetition period $\tau_r$ of 12.5 ns. The helicity of the pump pulses is modulated at a frequency of 50 kHz, thereby alternating between spin excitation along and against the $z$ direction, facilitating lock-in detection of the Kerr signal.

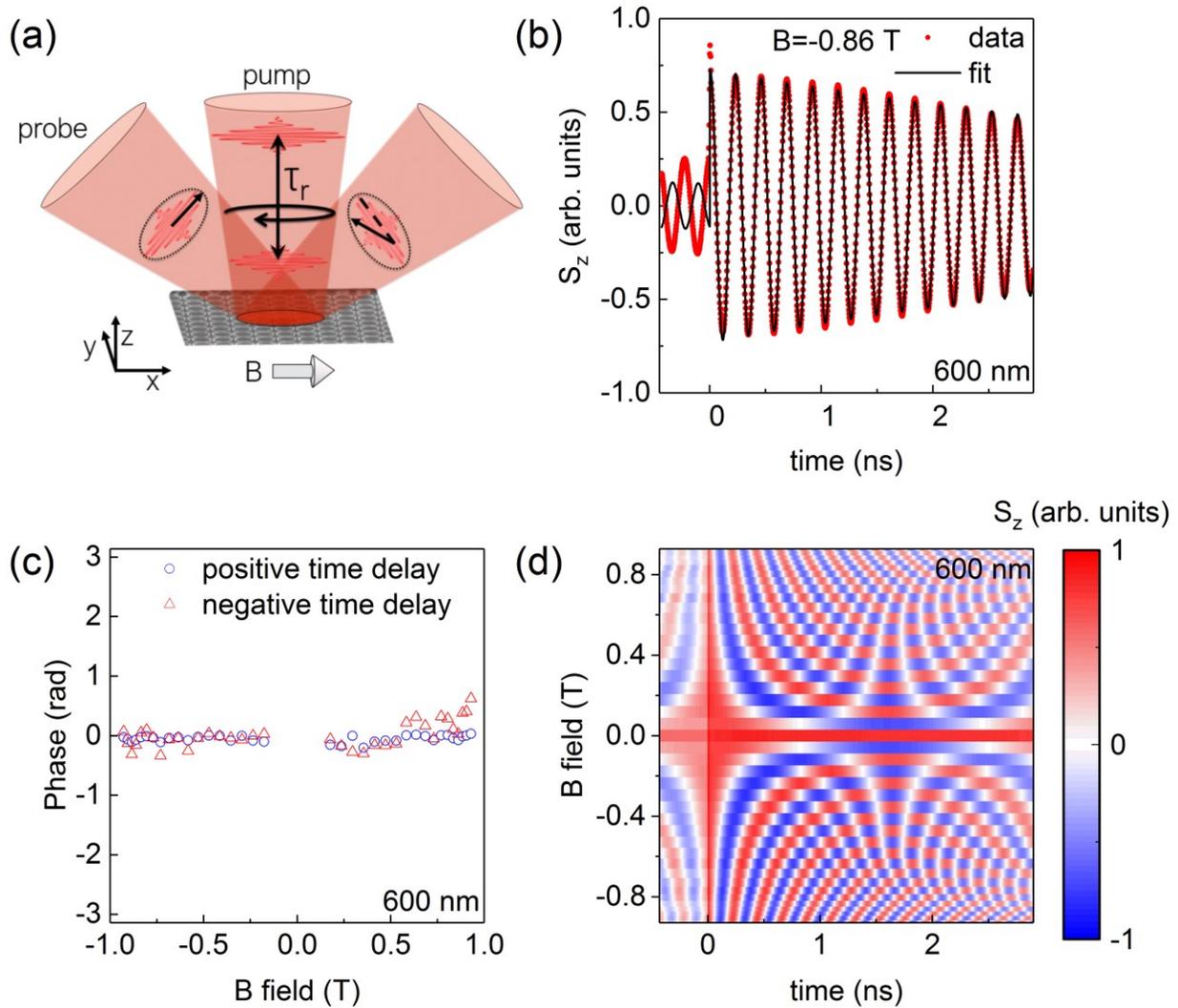

Figure 1: **Time-resolved spin dynamics in dot arrays**. (a) Sketch of time-resolved Kerr rotation measurement of spin component $S_z$. (b) Recorded $S_z(t)$ at a fixed external magnetic field of a 600 nm dot array. (c) Extracted phase of the oscillation $S_z(t)$ at positive zero-delay ($t_0^+$) and at negative zero-delay ($t_0^-$) of a 600 nm dot array. (d) Recorded $S_z(t)$ as a function of external magnetic field B for 600 nm dot size array. The values of $S_z(t)$ are color coded and normalized to +/- 1.

Figure 1(b) shows a time-resolved Kerr signal (red points) taken on an ensemble of 600 nm large dots and with $B$= -0.86 T. The signal is proportional to the electron spin component $S_z$. At zero-delay time $t$, a new pump pulse excites spin polarization that subsequently precesses in the external magnetic field, together with spin polarization that persists from the previous pump pulses. For increasing positive time delays, the precession amplitude decays. Such a decay is associated with two contributions. The first one is due to the electron spin lifetime ($\tau$) of a single dot and the second one arises from the inhomogeneous frequency broadening of the dot-ensemble ($\tau_{\text{inh}}$). An exponential fit of the precession amplitude [black line in Fig 1(b)] for $t$ between 0 and 2.8 ns yields an effective spin lifetime of $\tau^* = 6.8$ ns, significantly enhanced in comparison to the measured time in the unstructured two-dimensional electron gas (0.5 ns), but smaller than the laser repetition period $\tau_r$.

It has been observed in ensembles of self-assembled QDs [20] that the distribution of precession frequencies develops into a comb-like spectrum with a spacing given by the laser repetition rate. In the time domain, if $\tau > \tau_r > \tau_{inh}$, the spin polarization of such a spin-mode-locked system decays after a pump pulse but reemerges before a next pump pulse arrives, with identical spin precession phase immediately before ($t = t_0^-$) and after ($t = t_0^+$) the pump pulse arrival. The extrapolation of the fit in Fig. 1(b) to time delays between 12.07 and 12.5 ns ($t_0^-$) shows that such mode-locking must be present since there, $S_z$ is about twice larger than the fit, and the phase of spin precession is different. The phase of an unlocked system at a fixed $t$ should increase linearly with $B$, whereas it remains constant at $t = t_0^-$ in a mode-locked case. We have repeated measurements [see Fig. 1(b)] at varying $B$. Obtained results are shown in Fig. 1(d) as a 2D-map. The two axes represent $B$ and $t$, respectively. Over the full field range, we find a positive spin orientation at $t_0^-$. From fits of the Kerr signal at $t_0^+$ and $t_0^-$, we extract the respective spin precession phases. At $t_0^-$, we find a constant phase close to zero irrespective of $B$, clearly demonstrating spin mode-locking. We have omitted data points in Fig. 1(c) close to $B = 0$ since the determination of the phase is difficult if the spin precession period exceeds the available range of time delay. Spin mode-locking is consistently observed for all dot sizes between 400 and 1800 nm (see supplementary materials).

Spin mode-locking as previously observed in singly charged self-assembled QDs [19] has been explained by three different models. They all rely on the specific property of trion excitation where the spin polarization of the electron added by the pump pulse depends on the orientation of the resident electron spin before the pump pulse [29-30]. This is due to the Pauli principle that requires the excitation of an antiparallel spin configuration in the lowest-energy trion [31]. If $\tau_r$ equals an integer number of the spin precession period (phase synchronization condition - PSC), this mechanism leads to spin saturation effect (SSE) [20] where spin polarization is reduced as compared to the resonant spin amplification in a many-electron system [21, 32-33]. As has been shown in Ref. [21], this modification may lead to a spin mode-locked signal, but not because the distribution of spin precession frequencies is changed to a comb-like spectrum, but just because the spin polarization saturates differently for different precession frequencies. A further explanation [22] considers a time-dependent Knight field that originates from the average transverse electron spin polarization and induces nuclear magnetic resonance. The trion-related SSE reduces the Knight field at PSC, hence locking the nuclear polarization. The third model [20] considers optically stimulated fluctuations of the nuclear spin polarization. At PSC, optical excitation of trions is reduced, and because of an energy bottleneck from the different Zeeman energy of electrons and nuclear spins, nuclear fluctuations are strongly suppressed, thus keeping the dot-ensemble in a mode-locked condition.

In our QDs the Fermi energy is large enough such that excitation of trions is screened and band-to-band transitions dominate [34]. In such a scenario, additional spin polarization can always be accommodated irrespective of previous polarization, as long as the total spin polarization does not exceed 100%. This contrasts with the assumption in all previous explanations of spin mode-locking. In the following, we first investigate whether spin mode-locking has a purely electronic origin or whether it is related to nuclear spin polarization. We then develop a model that qualitatively explains our experimental observations.

**Nuclear origin of mode-locking**

We depolarize nuclear spins by leaving the sample in a zero external magnetic field and by blocking the laser pulses with a mechanical shutter. The system is kept in this condition for 3 minutes. We then ramp the magnetic field to target field $B$, open the shutter, and immediately start to record six scans where for each scan $t$ is swept from 50 to -430 ps within a laboratory time of 15 s. We repeat the same procedure for different $B$ and obtain $S_z(B,t)$ for different laboratory times. Measurements are shown in Fig. 2 for a dot ensemble (400 nm diameter) with 200 µW pump and 10 µW probe power. In the 1$^{st}$ scan [Fig. 2(a)], taken directly after nuclear spin depolarization, we observe that as a function of $B$, $S_z$ oscillates about zero at $t = -15$ ps (dashed line). This is consistent with an unlocked spin system where the spin precession phase is given by $\varphi = \omega \cdot t$, with $\omega$ the angular precession frequency. We extract $\omega$ from positive time delays and show the expected lines of maximum positive $S_z$ ($\varphi = 2\pi n$) as black lines in Fig 2(a). Down to -100 ps, the measured spin pattern indicates an essentially unlocked spin system. At more negative $t$, regions of constant $S_z$ start to deviate from the expected lines of constant phase and instead align into bands of constant t, an indication for mode locking. In the 6$^{th}$ scan [Fig. 2 (b)], starting almost 80 seconds after nuclear spin initialization, the mode locking is completed and is visible as a pure oscillation of $S_z$ in $t$ with a phase that is independent on $B$. The observed disappearance of spin mode-locking for depolarized nuclear spins indicates that the mechanism must originate from hyperfine interaction to nuclear spins. To extract the characteristic mode-locking build-up time $t_{ML}$ in our system, we analyze traces of $S_z$ taken at $t = -15$ ps [dashed lines in Fig. 2 (a) and (b)]. For an unlocked system we expect $S_z(B)$ to oscillate about a zero-offset value. With a focusing of the spin precession frequencies onto PSC modes, the oscillation amplitude decays and an offset should appear. Both the amplitude and the offset can be taken as a degree of spin mode-locking. The measured slices of $S_z(B)$ at $t = -15$ ps for all six scans are shown in Fig. 3(a). Already the first trace shows a small offset, indicating that the process of mode locking has already started. With increasing laboratory time (scan number), the oscillation amplitude decreases while the offset increases, hence the system is approaching the mode-locking state. A fit of the amplitude and offset is shown versus laboratory time in Figs. 3(b) and (c). In these fits, we take into account a linear decrease of the offset and oscillation amplitude with $B$, which we attribute to a $B$-dependent spin lifetime. From the laboratory time dependence of the offset and amplitude, we extract $t_{ML}$ to be in the order of 25 seconds, roughly independent of pump power [data for different pump power is shown in Figs 3(b) and (c)]. This is fully compatible with previous studies on DNP which report that the time scale of DNP is independent of pump power and takes place on seconds time scale [35]. We exclude that only statistical nuclear fluctuations as proposed in [20] are driving the mode locking because we observe a decay time of spin mode-locking that is on the same order as the build-up time (shown in the supplementary). This demonstrates that for our samples and measurement conditions nuclear polarization is not frozen at PSC by bleaching of the optical excitation. We want to point out that DNP leads to a frequency change of the average electron spins, which is visible in the phase relation of individual scans (scan 1 to scan 6) in Fig. 3 (a). The phase of $S_z(B)$ at a fixed pump-probe delay is proportional to $\cos((\omega + \omega_n) \cdot t) = \cos(\omega t + \varphi_n(t))$, so the change in phase can be attributed to a variation of $\varphi_n$ with laboratory time, induced by DNP.

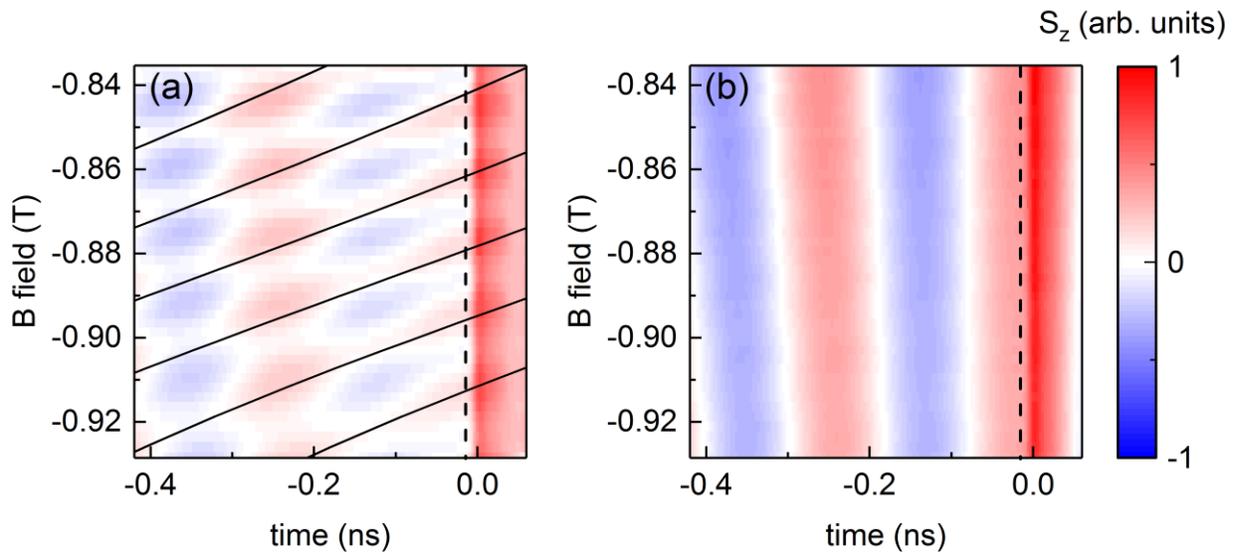

Figure 2: **Dependence on nuclear polarization**. (a) Recorded $S_z(t)$ of a dot array (400 nm) immediately after depolarization of nuclear spins. The magnetic field is scanned from -0.93 T to -0.83 T in 5 mT steps. Same normalization of $S_z$ is used in (a) and (b). The black lines mark the expected positions of constant phase where $S_z$ should attain maximum values in an unlocked spin system. (b) Signal from 6$^{th}$ scan, recorded 90 s after nuclear spins depolarization. The system is saturated and spin precession is mode-locked to the repetition rate of the laser, visible as a constant phase of the spin signal at $t = -15$ ps (dashed line).

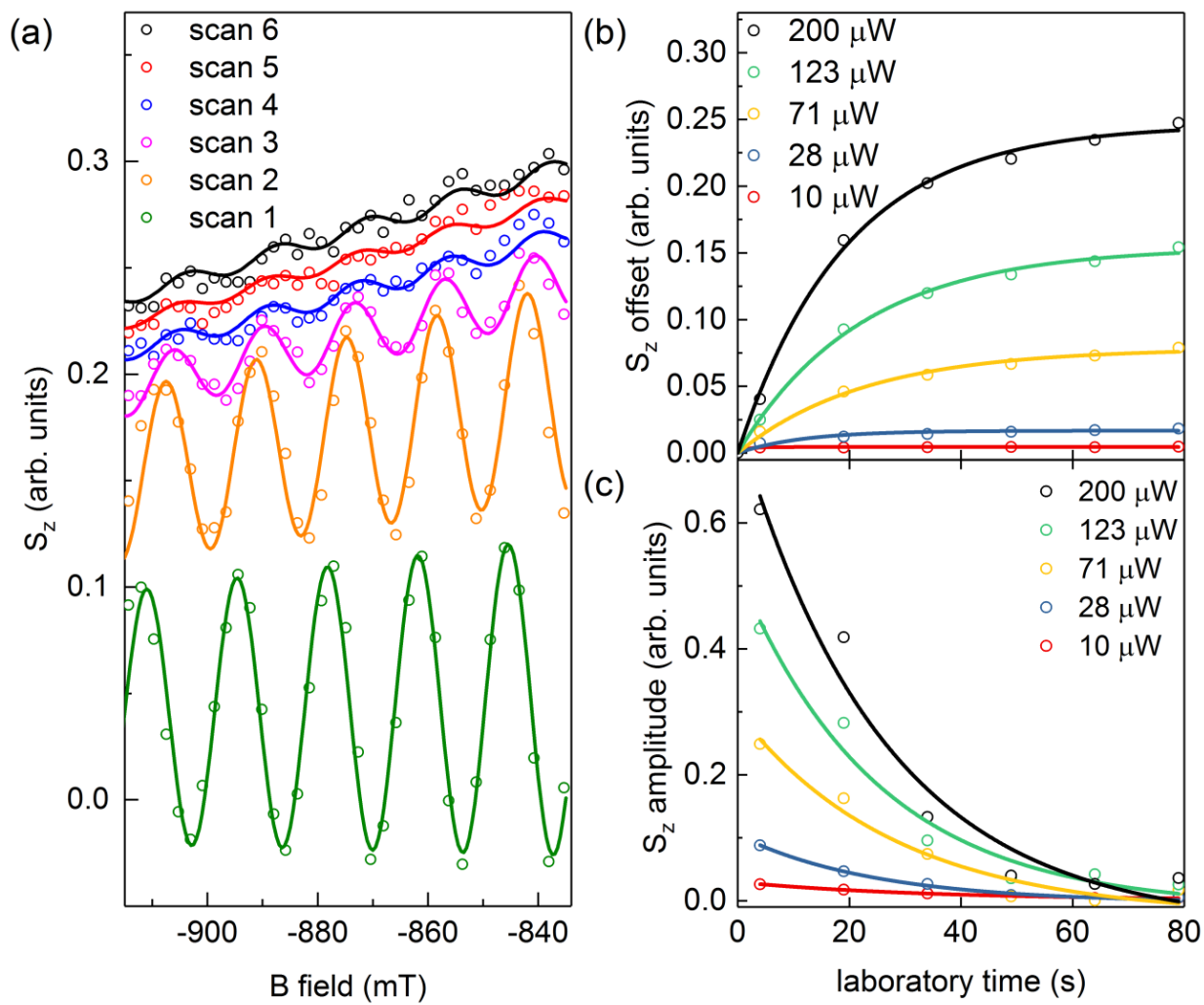

Figure 3: **Build-up of mode-locking**. (a) Magnetic field dependent Kerr signal at $t = -15$ ps for six scans taken at different times after depolarization of the nuclear spin polarization (difference between each scan is 15 s). An increase in offset or decrease in amplitude is a measure for mode-locking. (b)/(c) Extracted offset and amplitude of the oscillations in $B$ from the fit in (a). Data is shown for different pump powers and a fixed probe power of 10 µW. The characteristic offset buildup time or amplitude decay time is pump power independent and is extracted from the fit to be 25 s.

**Model for mode-locking mechanism**

The finding that mode-locking also occurs in many-electron QDs with a purely nuclear origin asks for a new explanation of the effect. We support our experimental results with a model that contains both DNP and SSE. First, we want to elaborate how SSE is considered. Similar to Ref. [21], we assume that the added spin polarization per pulse, $\Delta S = S_{new} - S_{old}$, depends on the spin polarization $S_{old}$ before the pulse. We fix the maximum spin polarization of the Fermi sea to a value $S_{max} = \pm\frac{1}{2}$ by assuming $\Delta S$ to saturate with an exponential factor:

$$\Delta S = P(1 - \exp(-|S_{max} - S_{old}|/P)), \tag{1}$$

whereby $P$ is the unsaturated spin excitation per pump pulse (see supplementary for a graphical representation). Starting with $S_{old} = 0$ and summing over enough laser pulses, we numerically determine the saturated spin polarization $S_{old}$ as a function of spin precession frequency. In Fig. 4(a), the obtained $S_{old}$ is compared to that from an unsaturated system (described by $\Delta S = P$), equivalent to the standard resonant spin amplification result [21,33]. The unsaturated curve exceeds the maximum spin polarization of $S_{max} = \frac{1}{2}$ at PSC. The saturated curve remains always below $S_{max}$, which goes parallel with a substantial dip in $\Delta S$ within a narrow range around PSC, see Fig. 4(b). As we will see in the following, this dip leads to a decrease in DNP which steers the nuclear spin polarization to a stable situation where electron spins mode-lock.

The hyperfine interaction is responsible for an Overhauser field $\mathbf{B}_n$ that is proportional to the nuclear spin polarization, and to DNP which changes $\mathbf{B}_n$ proportional to the average electron spin polarization $\langle \mathbf{S} \rangle$. This leads to a feedback mechanism that is described by the following coupled equations [35-36]:

$$\frac{d\mathbf{S}}{dt} = \frac{\mathbf{S} - \mathbf{S}_0}{\tau} - \frac{g\mu_B}{\hbar} \mathbf{S} \times (\mathbf{B} + \mathbf{B}_n) \tag{2}$$

$$\mathbf{B}_n = K \frac{\langle \mathbf{S} \rangle \cdot (\mathbf{B} + \alpha \langle \mathbf{S} \rangle)}{B^2} \mathbf{B} \tag{3}$$

The first term on the right side of eq. (2) describes the electron spin relaxation, where $\mathbf{S}_0$ is the steady-state spin. The second term describes the electron spin precession in the total magnetic field, $\mathbf{B} + \mathbf{B}_n$. The steady-state nuclear field $\mathbf{B}_n$ can be calculated by eq. (3), whereby $K$ is a material dependent constant and α is the Knight field constant. In the case of even a small misalignment of the magnetic field direction to the sample plane, $\mathbf{S}_{new} - \mathbf{S}_{old}$ has a finite component along **B**, leading to a DNP drive component $\langle \mathbf{S} \rangle \cdot \mathbf{B}$ that is proportional to $\Delta S\, \tau/\tau_r$ [see sketch in Fig. 4(c)]. The transverse components of $\langle \mathbf{S} \rangle$ enter the drive term proportional to $\alpha \langle \mathbf{S} \rangle^2$ in Eq. (3), which can be neglected due to the small α. We numerically solve the coupled system by taking the SSE into account (for detailed discussion see method section). The simulation result in Fig. 4 (c) shows the time dependent DNP drive component $\langle \mathbf{S} \rangle \cdot \mathbf{B}$ for a spin at an initial frequency (at $\mathbf{B}_n = 0$) of 4.30 GHz. It goes through two minima and after a certain time approaches a plateau. To understand such a behavior, we perform calculations for a set of initial start frequencies, express the total frequencies in terms of the laser repetition rate, and plot them versus laboratory time in Fig. 4(d). The time-dependent DNP drive in Fig. 4(c) corresponds to the red curve in Fig. 4(d). We find that the minima in the DNP drive term coincide with PSC, directly related to the narrow dips in $\Delta S$ at PSC. These reduction of the DNP drive at PSC realign the initially uniformly distributed electron spin precession frequencies into discrete PSC modes. This corresponds to the observed spin mode-locking and occurs on a timescale given by the DNP process (evaluated as $t_{\text{ML}}$ in the experiment). The spectral width of the dips in $\Delta S$ and therefore also of the PSC modes is proportional to the spin decay rate $1/\tau$. Importantly, this decay rate is sampled over the build-up time of spin polarization by the laser pulse train, i.e. at most a few $\tau$. Similar to dynamic decoupling protocols and other methods [37], but here in a self-driven way, this mechanism acts as a high-pass filter for nuclear fluctuations with a cut-off frequency at $1/\tau$, leading to a stabilization of nuclear polarization and a corresponding increase in electron spin lifetime.

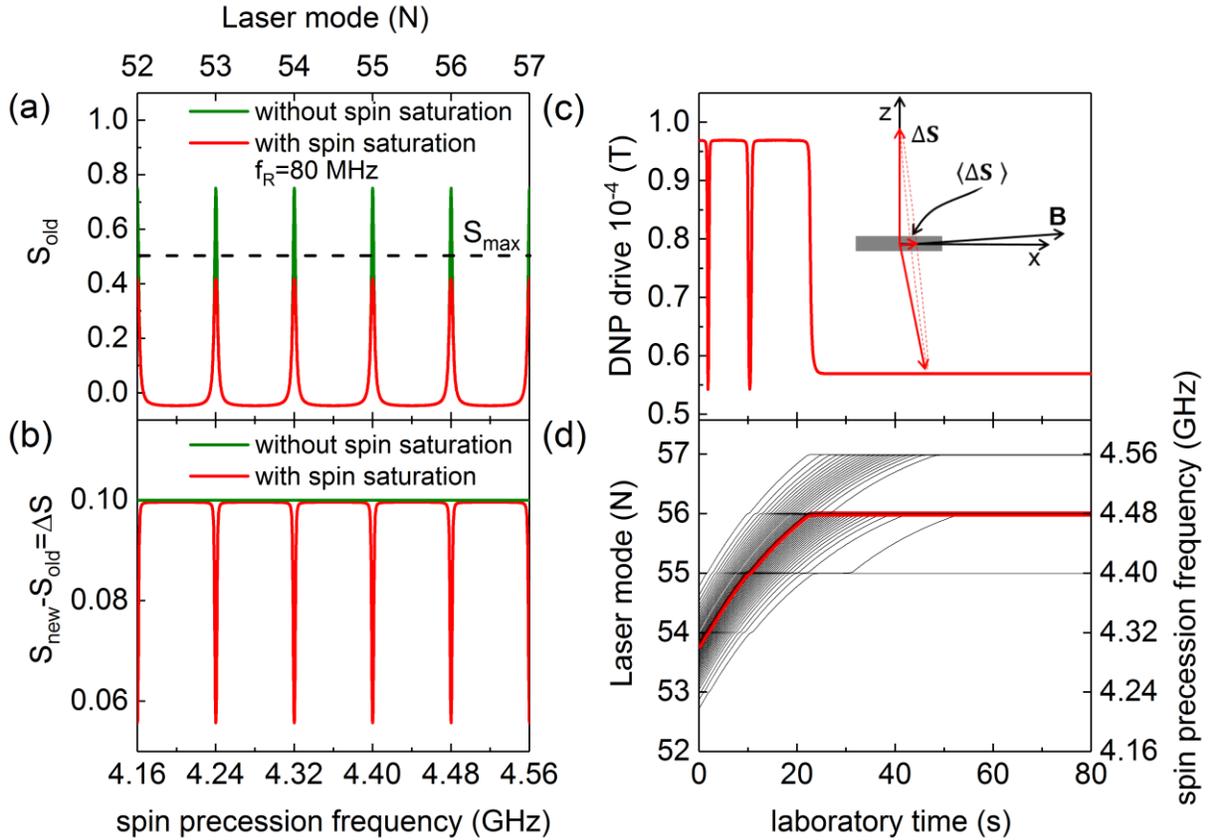

Figure 4: **Model for spin mode-locking.** (a) Calculated resonant spin amplification for an unsaturated spin system (green) and a saturated system (red). (b) Calculated spin excitation per laser pulse using eq. (1) and assuming a spin polarization per pulse of *P*=0.1 for saturated (red) and unsaturated (green) spin system. (c) Dynamical nuclear polarization drive $\langle S \rangle B$ calculated from the coupled equations (2) and (3). The insert shows the added spin polarization per pulse, $\Delta S = S_{new} - S_{old}$, with a non-zero average along **B**. (d) Black lines show evolution of the spin precession frequency for a distribution of initial start frequencies centered at 4.30 GHz (between 52$^{nd}$ and 55$^{th}$ laser repetition mode). Evolution of red marked frequency corresponds to the DNP drive shown in (c).

**Summary and outlook**

We have shown that the precession of localized electron spins in nano- and micrometer-sized structures can be synchronized to a periodic drive. This synchronization occurs because of a rearrangement of nuclear spin polarization in each QD. By a hyperfine-induced feed-back mechanism, the additional precession frequency from the nuclear polarization is focused to a width $\sim 1/\tau$. In this way, the electron coherence time in principle can overcome the limit given by a fluctuating nuclear background. This approach can be applied in general to all materials with nuclear background and a saturable spin system, thus in addition opening possibilities to synchronize electro-optical and opto-mechanical systems to localized spins. Here demonstrated using an optical pulse train, spin mode-locking can in principle be driven purely electrically or via piezoelectric effects also mechanically. The observed long electron spin lifetimes in many-electron QDs may also enable to couple such spins to microwave and phononic cavities, exploring the regime of cavity-quantum electrodynamics with localized spins [38-40].

**Methods**

**Sample fabrication**

The sample is fabricated from a modulation-doped GaAs/AlGaAs quantum well structure which is grown by molecular beam epitaxy. The quantum well is 18 nm wide and is located 143 nm below the sample surface. The carrier density of the quantum well is $n_{\text{2D}} = 2.15 \cdot 10^{11}$ cm$^{-2}$ under illumination. For fabrication of a QD ensemble we use positive resist for electron beam lithography. After resist developing, 150 nm of aluminum is evaporated. The aluminum layer serves in the next step as an etch mask for reactive ion etching with a HBr plasma. In this way the material around the deposited aluminum is removed (700 nm is etched from the surface). In the last step, the deposited aluminum is removed with KOH solution.

**Experimental settings**

The system is investigated with degenerated pump-probe spectroscopy with ps laser pulses at a wavelength of 811.46 nm which is in resonance with the quantum well transition. All experiments are performed at 15 K. The laser repetition rate is 80 MHz. The focused laser spot size of pump and probe is 30 μm. The probe power is fixed for all experiments and is 10 μW.

**Data fitting**

The Kerr signal in Fig. 1 (b) is fitted with an oscillating exponential decay of the form $S_z(t) = A \cdot e^{-\left(\frac{t}{\tau}\right)} \cdot \cos(\omega t + \varphi)$ for positive time delays. A lifetime $\tau^*$ of 6.8 nm is obtained for the data shown in Fig. 1(b). The same fitting function was used to extract the phase information ($\varphi$) in the data shown as a 2D map in Fig. 1(d). The fitting function is applied to a short time delay range at positive and negative time delays for each magnetic field $B$, and the extracted phase is shown in Fig. 1(c).

**Spin dynamics model**

Eq. (2) can be solved analytically, and the solution is decomposed into two orthogonal components transverse to **B**. Both components are defined by exponentially decaying harmonic oscillations. Spin accumulation is calculated by propagating $\mathbf{S}_{new}$ to $\mathbf{S}_{old}$ over a time $\tau_r$ using Eq. (2) and then determining a new $\mathbf{S}_{new} = \mathbf{S}_{old} + \Delta S \hat{\mathbf{z}}$ until a steady-state spin polarization is obtained. In case of no SSE, the resonant spin amplification equations are recovered [21,33]. In case of even a small misalignment of the magnetic field direction to the sample plane, the additional spin polarization $\Delta S\hat{\mathbf{z}}$ has a finite component along **B**, leading to DNP. Such DNP is reduced by employing a modulation scheme for the helicity of the optical pump pulses, but in practice is not cancelled to zero. In the simulation, we assume a misalignment angle of 2 degrees and an imbalance between left- and right-circular polarization of 5%. From eq. (2) we obtain ⟨**S**⟩, from which the target nuclear field defined in Eq. (3) is calculated. The actual nuclear field $\mathbf{B}_n$ is adapted in small steps towards the target field with the long time constant of DNP, in each step solving again equation (2) and calculating a new target field. We want to point out that we assume $\mathbf{B}_n$ to be aligned along **B**, which is a good approximation considering the small Knight field in GaAs. We use the parameters $gK$=7.79 T [35] and $\alpha$=0.

**Author contributions**
S.M. carried out the experiment and analyzed the data in close collaboration with G.S. Sample growth was performed by C.R. and W.W. and processed by S.M. The model was developed by G.S. All authors discussed the results in full details. S.M. and G.S. wrote the manuscript.

**Acknowledgements**
We acknowledge financial support from the NCCR QSIT of the Swiss National Science Foundation. We also thank R. Allenspach for helpful discussions, A. Olziersky and U. Drechsler for technical assistance.

**Competing interests**
The authors declare no competing interests.

# Supplementary materials for universal nuclear focusing of confined electron spins

**Estimation of electron numbers in a single quantum dot**

The number of electros in each quantum dot is given by the dot size and the carrier density of the quantum well. From the Hall measurements we extract the carrier density of the quantum well to be $n_{2D} = 2.15 \cdot 10^{11}\ cm^{-2}$ under illumination. For our estimation we neglect the depletion length ($l_D$) of the carriers at the edge of the quantum dot, assuming that the quantum dot diameter is much larger than $l_D$. The table below shows the calculated number of electrons in each dot.

| Dot diameter (nm) | Number of electrons |
|---|---|
| 1800 | 5478 |
| 1000 | 1690 |
| 600 | 608 |
| 400 | 270 |

*Table 1 :* **Dot carrier density**. Calculated number of electrons in a single quantum dot from the quantum well carrier density $n_{2D}$

**Mode-locking on different dot diameters.**

In addition to the data presented in the manuscript we show that spin mode-locking occurs for different dot diameters. Exemplary we show in Fig. 1 (a) a scanning electron microscopy image of a 600 nm dot array. In Fig. 1 (b)-(c) data on spin mode-locking is presented for dot diameters of 500, 1000 and 1800 nm. The data in Fig.1 (b)-(c) is recorded in the same way as described in the manuscript. For the 1800 nm dot array we observe that at high *B* and negative time delays the spin signal becomes weak. Such a behavior is attributed to a reduced spin lifetime induced by the larger influence of spin-orbit dephasing, and to a magnetic-field dependent spin lifetime.

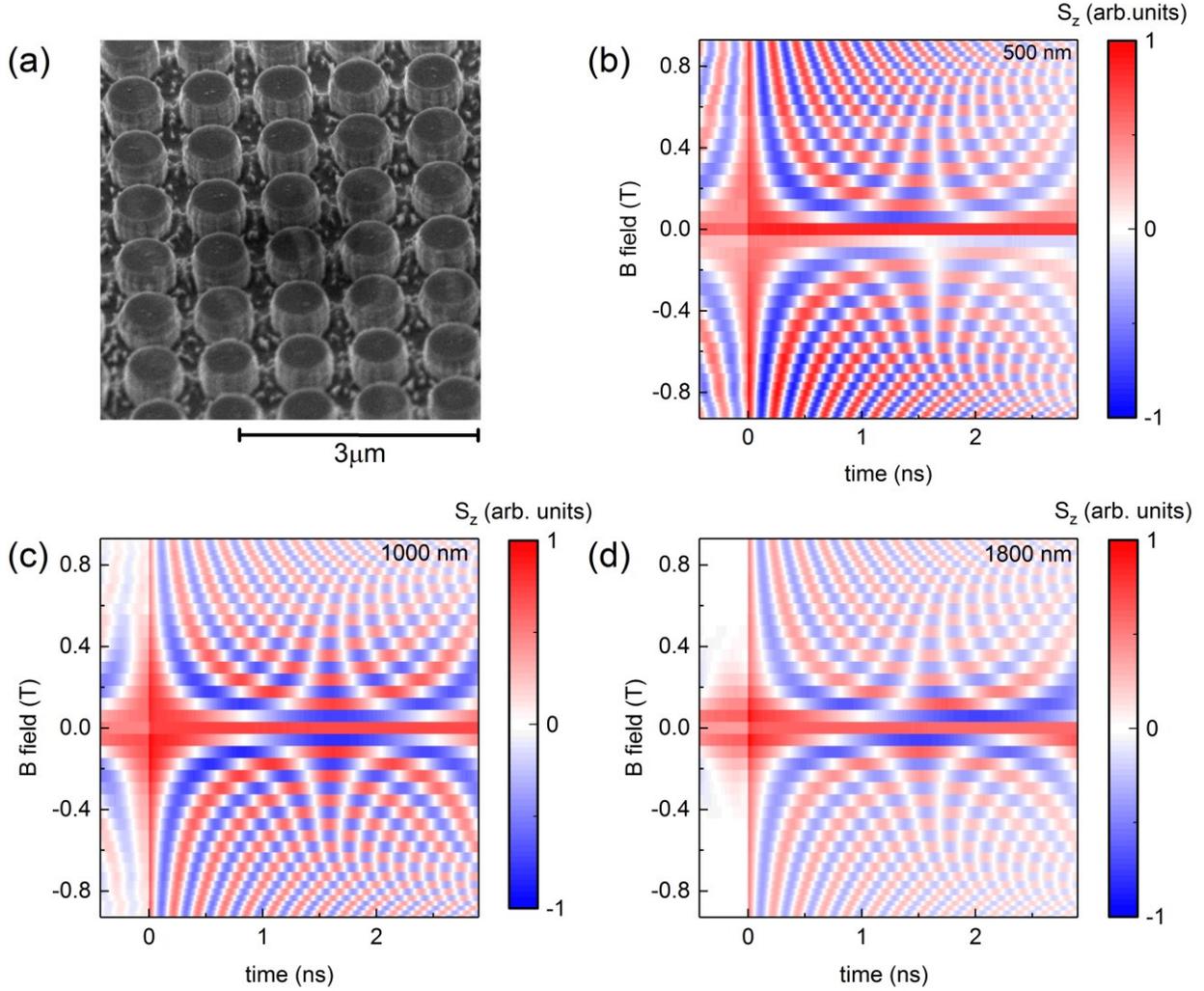

Figure 2: **Spin mode-locking for different dot diameters**. (a) Scanning electron micrograph of lithographically defined quantum dots. (b)-(d) Time-resolved Kerr signals as a function of external magnetic field $B$ for 500, 1000 and 1800 nm dot size arrays. The $S_z$ signal is color coded and is normalized to +/- 1.

**Model for spin saturation effect (SSE)**

In order to account for the SSE, we define a spin dependent spin excitation function $S_{\text{new}}(S_{\text{old}})$. Equation (1) in the main text describes a heuristic function and is plotted in Fig. 2 (red curve) for an additional spin polarization per pulse of $P = 0.1$. For comparison, the saturation curve for singly-charged quantum dots with trion excitation is shown as a blue curve [1]. For trion excitation, the additional spin excitation per pulse depends linearly on $S_{\text{old}}$. In our model for a Fermi sea, it is constant unless $S_{\text{old}} + P$ becomes close to full polarization.

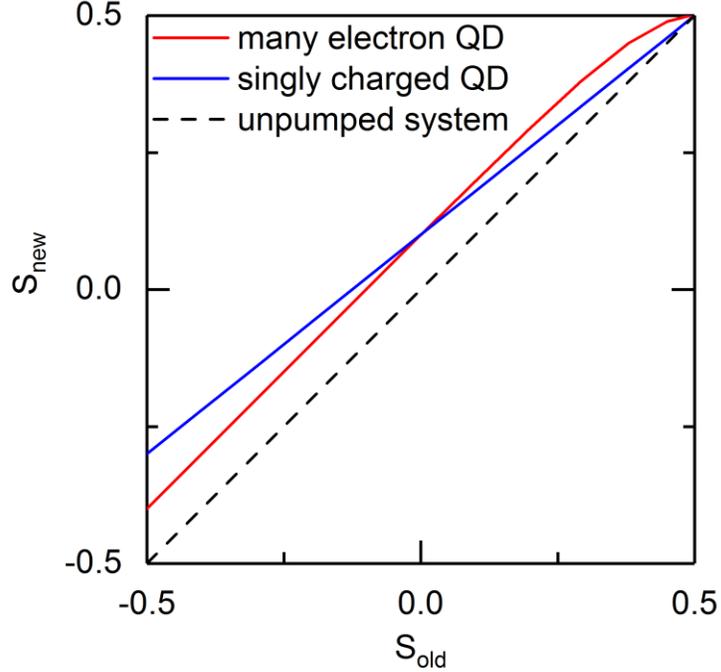

Figure 2: **Spin saturation**. Spin dependent spin excitation for singly charged quantum dots and for multiple charged quantum dots (red curve) as used in our model (spin polarization per pulse is $P = 0.1$).

**Decay of spin mode-locking**

We describe measurements of the characteristic decay time of spin mode-locking after the periodic spin excitation is switched off. For this, we first saturate the mode-locking by periodical optical excitation for 5 min in an applied external B field of -0.93 T. We then block the laser pulses with a mechanical shutter, wait for a certain time and after unblocking the laser record the Kerr signal for time delays between -0.43 ns to 2.8 ns. We repeat this procedure for waiting times $t_{\text{lab}}$ (laboratory times) between 5 and 150 s. The obtained data is shown in Fig. 3 (a). For negative time delays, we observe a change in the Kerr signal phase with increasing laboratory time. Additionally, we observe a decrease of the Kerr signal amplitude at negative delay times. Both are due to depolarization of the nuclear spins, resulting in a decrease of the nuclear field from the value obtained by DNP, and in a broadening of the ensemble spin precession frequencies. This is in agreement with our model (see Fig. 4 (d) in the manuscript), showing that the Larmor precession frequencies are not only distributed into discrete modes, but their average also increases as compared to the depolarized case. We see the change in the average precession frequency also from the Kerr signal at positive time delays. Figure 3 (b) shows measured $S_z(t)$ at laboratory times 5 and 140 s. The two signals oscillate at different frequencies, which we extract by fitting oscillating decaying curves for positive time delays. The frequency in the mode-locked case (5 s laboratory time) is 22 MHz higher than in the partially locked case (140 s laboratory time). While taking the measurement of $S_z(t)$, the laser pulses drive spin mode-locking, thus the fitted Larmor precession frequency at 140 s is slightly overestimated. A dynamical frequency shift is also reflected in our model [Fig. 4 (d)]. The initial Larmor precession frequencies (without DNP at laboratory time 0 s) shift to a higher frequency once the DNP sets in.

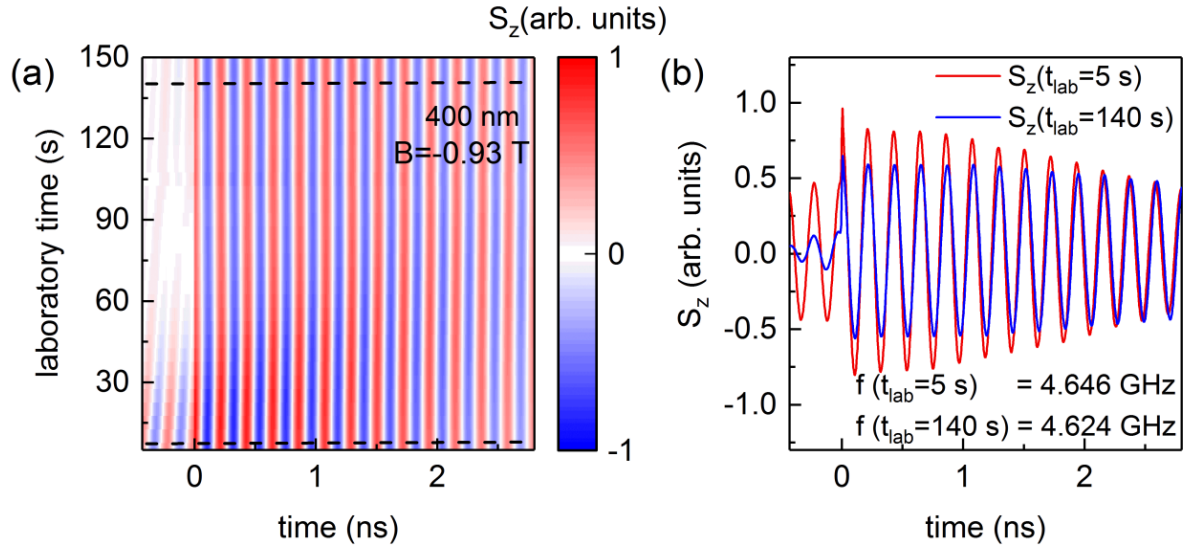

Figure 3: **Mode-locking decay**. (a) Decay of spin mode-locking studied in 400 nm dot array at B=-0.93T. The Kerr signal (measuring the spin component $S_z$) is color coded and is normalized to +/- 1. (b) Slices of (a) which are indicated with dashed lines. Kerr signal at laboratory time 5 s (red curve) and 140 s (blue curve) exhibit different precession frequencies at positive time delays, indicative for a decay of the DNP.